\documentclass[twocolumn]{aa}
\usepackage{graphicx}
\begin{document}
  \title{On the Outbursts of Black Hole Soft X-Ray Transients} 

   \author{ {\"U}nal Ertan \inst{1} 
            \and  M.Ali Alpar \inst{2}}

   \offprints{{\"U}nal Ertan}

   \institute{Physics Department, Middle East Technical University 
                Ankara 06531, Turkey \\  
     \email{unal@astroa.physics.metu.edu.tr}
\and 
      Sabanc{\i} University, Orhanl{\i}-Tuzla 
         81474 {\.I}stanbul/ Turkey  \\
\email{alpar@sabanciuniv.edu.tr}
                               }
   \date{Received March 4, 2002; accepted July 4, 2002}

%\leftline{\bf{Thesaurus code numbers:} 02.01.2, 08.14.2, 13.25.5 }\\
%{\bf{Section : Accretion, accretion disks - novae - X-rays:
%stars}  
%~~~  } \\ \centerline{ ~~~ }\\
%\leftline{ Proofs to: {\"U}.Ertan}\\
%\leftline{ }}

\abstract{

We suggest a new scenario to explain the outburst 
light curves of black hole soft X-ray transients
together with the secondary maximum and the bump seen on their 
decay phases. Our explanations are based on
the disk instability models considering  the effect of X-ray
irradiation. The scenario is consistent with the observed X-ray delays
by a few days  with respect to the optical rise for both the main 
outburst and the later maxima. We  test our scenario by  numerically
solving 
the disk diffusion equation.  
The obtained model curve fits well to the observed X-ray 
outburst photon flux curve 
of the black hole soft X-ray transient GS/GRS 1124-68, 
a typical representative of the four BH SXTs with very similar 
light curves. 
  }

\def\la{\raise.5ex\hbox{$<$}\kern-.8em\lower 1mm\hbox{$\sim$}}
\def\ga{\raise.5ex\hbox{$>$}\kern-.8em\lower 1mm\hbox{$\sim$}}
\def\ea{\it et al. \rm}
\def\am{$^{\prime}$\ }
\def\as{$^{\prime\prime}$\ }
\def\eg{{\sl EGRET }}
\def\be{\begin{equation}}
\def\ee{\end{equation}}
\def\ba{\begin{eqnarray}}
\def\ea{\end{eqnarray}}
\def\d{\partial}
\def\R{\right}
\def\L{\left}
\def\D{\mit\Delta}
\def\S{\Sigma}
\def\bc{\begin{center}}
\def\ec{\end{center}}
\def\m{\mbox}
\def\o{\odot} 
\def\a{\alpha}
\def\Tirr{T_{\m{irr}}}
\def\Teff{T_{\m{eff}}}
\def\Hirr{H_{\m{irr}}}
\def\Mdot{\dot{M}}
\def\Msun{M_{\odot}}
\def\xr#1{\parindent=0.0cm\hangindent=1cm\hangafter=1\indent#1\par}
%}

     \maketitle

\section{Introduction}

Soft X-ray transients (SXTs), a subclass of low mass X-ray
binaries, 
contain either a neutron star (NS SXTs) or a black hole (BH SXTs). 
Their sporadic outbursts with observed 
or estimated recurrence time 
scales changing from months to more than 50 years show a variety of light 
curves (Chen et al. 1997). Among these sources the BH SXTs    
GRO~J0422+32, A0620-00, GS/GRS~1124-68 and GS~2000+25 show very similar 
X-ray  outburst light curves. Rise to maximum is fast (few days). 
Decay can be fitted with exponentials, or with power laws.
These four sources are usually labeled as FRED
(fast-rise-exponential-decay) 
sources. We refrain from using the term FRED because the initial decays
can also be fitted with power laws. 
Instead, we shall use the term   BH SXTs to cover only these 
four sources and the term ``exponential-like'' to describe their
similar secular decay behaviors. 
Other common features of the outburst light curves are  a typical 
secondary maximum 
 seen in the decay phase about two months after the main maximum  
and a bump seen at the end of the 
decay phase, after the secondary maximum. 
The secondary maximum is exhibited 
by all four BH SXTs above. The bump is 
seen in the outburst light curves of A 0620-00 and  GS/GRS~1124-68.    
It is not very clear in the light curve of GS~2000+25 and absent in that 
of GRO~J0422+32 (Tanaka $\&$ Shibazaki 1996).

Three physical effects are likely to determine the behavior of SXTs. 
These effects are (i) viscous evolution of the accretion disk, (ii)
transitions between the two stable states of the disk through hydrogen
ionization, as in disk instability models, and (iii) irradiation of the
disk by X-rays from the inner disk (also from the neutron star surface in
the case of NS SXTs). In this paper we present a comprehensive model for 
the black hole SXTs based on an interplay between all three effects. 

The disk instability models (DIMs) which are successful in reproducing 
the general characteristics of dwarf nova (DN) light curves 
(Osaki 1974, Hoshi 1979, Meyer $\&$ Meyer-Hofmeister 1981) are also
suggested 
to be the possible mechanism for the SXT outbursts 
(Cannizzo et al. 1985, Huang $\&$ Wheeler 1989, 
Mineshige $\&$ Wheeler 1989).  
According to DIMs the equilibrium solutions on the accretion rate
$\dot{M}$ 
- surface density $\Sigma$ plane form an "S" shaped curve for 
a given radial distance R from the center of the disk. The upper and 
lower branches of the curve are viscously stable, 
whereas the middle branch is unstable. 
The unstable branch appears due to partial ionization of hydrogen and
corresponds roughly to the temperature range 
$\Teff \sim 6\times 10^3 - 10^4$ K. According to DIMs all the disk is in
the 
cold stable branch during the  
quiescent state. In the cold stable regime, 
surface densities increase with time at all radii due to the low 
accretion efficiency. The critical maximum
surface density of the cold 
branch is eventually exceeded first at some particular radius R, and 
the instability propagates as a density wave  
towards the smaller and the larger radii. If all or most of the disk 
jumps to the hot stable branch the increasing accretion rate leads to 
an outburst. For the disk to return to the quiescent phase the surface 
density should decrease below the critical minimum surface density of 
the hot stable branch at some location in the disk. This is expected to 
occur first at the outermost regions of the disk, where in the hot state 
the surface densities are lowest, while the critical surface densities are 
highest. 
Consequently, all the disk returns back to the cold stable regime by 
means of a cooling front propagating from the outer disk towards the 
inner regions. The observed decay in luminosities subsequent to a
burst is expected to 
be driven by the propagation of this cooling front. 

The standard version of DIMs have difficulties in explaining the  
long recurrence times and the slow decay characteristic 
of the SXT outbursts with plausible standard thin disk $\a$ parameters 
(Shakura $\&$ Sunyaev 1973) estimated  from DN light curve analyses.
These difficulties do not exclude the DIMs as the mechanism for the 
SXT outbursts, but require a modification of their 
standard form. Including the evaporation and the consequent disk
truncation in the models it is possible to explain the long recurrence
time scales (few decades or more) of BH SXTs (Meyer-Hofmeister $\&$ 
Meyer 1999, Dubus et al. 2001).  X-ray irradiation 
of the outer disk and  possibly the evaporation also seem to be 
responsible for the observed characteristics of the decay curves of BH
SXTs. 
It has long been known that the X-ray irradiation is the dominant 
source of the observed optical light of persistent LMXBs and in 
particular SXTs in their outburst states. It was suggested that 
the X-ray irradiation 
may have a stabilizing effect on the outer mass flow keeping the 
outer disk temperatures above the hydrogen ionization temperature 
(Meyer $\&$ Meyer-Hofmeister 1984, van Paradijs $\&$ McClintock 1994). 
In a previous work, we showed that 
pure viscous evolution of matter located originally at the outer disk 
can reproduce the rise, the turnover and the early decay characteristics 
of the BH SXTs by fitting the model light curves to the
observed X-ray photon flux data of the BH SXTs 
GS/GRS 1124-68 and GS 2000+25 (Ertan $\&$ Alpar 1998).   
This also suggests the importance of the X-ray
irradiation, since the irradiation  could build up the
conditions for a pure viscous decay by preventing the 
propagation of a cooling front for long times. 
By analytical treatments, King $\&$ Ritter (1998) (KR) concluded that 
the slow decay is produced when 
the disk is fully ionized by irradiation, for inner accretion rates 
above a certain critical $\Mdot_{\m{in}}$ 
and the decay seemingly linear in time that is seen  
at the end of the decay phase is produced 
below this $\Mdot_{\m{in}}$, when a cooling front propagates inwards 
from the outer disk.  
Detailed numerical calculations show that a disk fully 
ionized by strong irradiation  
gives a much slower decay than the observed
decays (Cannizzo 2000).  
By adjusting the evaporation rate in his model,  Cannizzo (2000)
obtained the observed decay behavior for a fully ionized disk. 
On the other hand, following the idea of KR, but employing a moderate
irradiation in models, it is also possible to reproduce the observed
decays with a negligible effect of evaporation on the decay curves 
(Dubus et al. 2001; DHL).    
 
The vertical structure analyses of an X-ray 
irradiated disk show that the direct illumination of the outer 
disk by the central X-rays is not possible due to the self screening of
the disk (Dubus et al. 1999).                
On account of the fact that the observations invariably give much 
higher $L_{\m{opt}}/L_{\m{x}}$ ratio than expected from intrinsic 
dissipation alone, Dubus et al. (1999) suggested that the X-ray
illumination is very probably still present, but 
indirect or the outer disk 
is warped. Indirect irradiation of the outer disk 
may take place via 
scattering. Even if the inner disk is vertically 
optically thick and geometrically thin, 
the surface layers where heating dominates over cooling 
could be 
thermally unstable and could evaporate to a hot corona  (Shaviv $\&$ 
Wehrse 1986). Existence of such a hot corona 
surrounding the inner disk  will be our main  
assumption regarding the mechanism of indirect irradiation. 

Since the optical flux is X-ray irradiation dominated,  
a change in the intrinsic dissipation at the outer disk can hardly be 
the reason for the observed enhancement in optical flux at the secondary
maximum and the bump. Irradiation must be incorporated in
the explanation of these features. In an alternative scenario  
tidal instabilities due to the 3:1 resonance 
(Frank et al. 1992 ) at the outer disk could  
modify the optical light of the BH SXTs, 
but only if the disk size increases considerably,  
intercepting more of the X-rays coming from the inner disk (Haswell et al. 
2000). However, it is unlikely that the consequent changes in the 
mass inflow rate in the outer disk 
can lead  to the observed secondary maximum or the bump. 
 
A further restriction on models comes from the comparison of
the optical and X-ray  
observations. Detailed analyses show that whenever it is possible to make 
a clear comparison between the optical and the X-ray observations of the 
main and the minor
 maxima of BH SXTs, it is seen that the optical rise 
precedes the X-ray rise. The X-ray delays with respect to the optical
light are $\sim$ 4-6 days for the main and the secondary  
maximum and around two weeks for the bump  
(Kuulkers 1998, Ebisawa et al. 1994, Orosz et al. 1997).   
Considering these observational constraints,   
we suggest an explanation for the overall outburst phase of 
BH SXTs including the secondary maximum (Sect. 2) and the 
bump (Sect. 3). 
The details of the numerical model
testing these explanations is  presented in Sect. 4.
In Sect. 5 we summarize and discuss the results.     

\section{The Main Outburst and The Secondary Maximum}

The secondary maximum  
is seen roughly two months after the main maximum. An abrupt 
increase in the X-ray light curve by a factor $\sim 1.5-2$~ is followed by 
a curve that remains roughly parallel to the extrapolation from the main 
decay in the logarithmic plot (Tanaka $\&$ Shibazaki 1996). 

Mineshige (1994) pointed out that an abrupt heating of the disk and a 
substantial mass
supply are required to account for the observed secondary maximum, 
and attributed the
onset of the X-ray heating of the disk to removal of a Compton cloud, which 
had been blocking 
the central X-rays from reaching 
the outer parts. He suggested that the mass supply is either due to
an enhanced mass transfer from the companion or a transient 
recession of the cooling front
in the disk. Some difficulties arise in either case. 
Mass transfer from the 
companion is unlikely, since the $L_1$ point is shielded by the disk. 
Even if the $L_1$ point is irradiated   
it is extremely difficult to increase the mass transfer rate by
 the X-ray heating unless 
the X-ray spectrum is very hard (King 1989). 
On the other hand , if X-ray irradiation were 
blocked by a central corona, then a cooling front would have propagated  
in the two months before the occurrence of the secondary maximum. 
In that case the optical luminosity would decay 
faster than the X-ray luminosity which is not the case. 

Chen et al. (1993) explained both the secondary maximum 
and the bump by an enhanced 
mass transfer rate from the companion star. 
They suggested that the secondary maximum is 
due to the heating of the outer layers of the atmosphere of the secondary 
when the outer 
disk becomes optically thin while the bump occurs 
via a mass transfer instability caused by hard X-ray heating of the
 subphotospheric layers of the  
secondary during the main outburst. 
They attributed the delay between the secondary maximum and the bump   
to the transport of part of 
absorbed hard X-ray energy to the entire 
convective region of the secondary by convection. 
It seems hard to account for  the abrupt rise of the mini 
outbursts by these mechanisms. In addition, if the $L_1$ point is 
indeed illuminated because of 
the evaporation of the outer disk then X-ray eclipses would be observed
in sources with moderate to 
high inclination. X-ray eclipses were not  seen 
during the decay phase of the secondary maximum from any of our four
sources. 

Since the observed optical light is likely to be
X-ray irradiation dominated, 
the optical rise during the secondary maximum  
is expected to occur due to either an increase in the 
central X-ray luminosity or a change in the outer disk properties, thereby
intercepting a larger fraction of the X-rays. Since the observed optical 
rise precedes the X-ray rise by a few days, the former possibility is not 
favorable. 
At the beginning of the main outburst, 
the accumulated matter at the outer 
disk expands to both smaller and larger radii. 
The effective X-ray heating of the outer disk 
cannot start immediately upon the triggering of the outburst. 
This takes a viscous time scale corresponding  to the expansion of the
released matter from the outer to the inner disk. 
In this time the outer disk may already have gone into 
the cold stable state by means of a cooling front propagation 
before the central X-ray luminosity  affects the stability of the 
outer disk. KR suggested that the X-ray irradiation 
can cause this outermost cold disk to jump back to the hot stable branch
at the very beginning of the outburst 
and increase the amplitude of the 
X-ray luminosity about $\sim 40-120$ days after the main X-ray maximum. 
In addition, they pointed out that the optical flux cannot rise until 
after the increased mass flow reaches the inner disk and increases the
X-ray radiation.   
This is contradicted by the observations. 
This suggestion was not confirmed by
detailed numerical model calculations (Cannizzo 1998, Dubus et al.
2001). By comparing the results of nonirradiated and irradiated 
disk models with the observations, Cannizzo (1998) concluded that 
a completely and strongly irradiated disk model overestimates 
the optical flux at the maximum light of A0620-00, and that the
irradiation temperatures 
$\Tirr\simeq 0.3-0.4 \Teff$ at the outer disk are enough to explain the
observed optical flux  
if all the disk is irradiated. Incorporating effects of evaporation 
in fully irradiated and completely ionized disk models,   
Cannizzo (2000) showed  that the evacuation 
of the inner disk due to the strong evaporation, and the subsequent
refilling could reproduce the observed 
features of both the main decay and the secondary maximum. Note, however,
that this model does not address the observed X-ray delays.

We now turn to the second option, a change in the outer disk properties. 
We propose that the part of the outer disk that made the 
transition to the  cold state before the increase 
of the central accretion rate  
associated with the main outburst may not be heated by the central 
X-rays for a while after the main outburst,  
even when the X-ray luminosity reaches its maximum. 
This is  because  the pressure scale height of 
the outermost  cold disk should be 
lower than that of the intermediate disk, at radii less than and 
of the order of a ``hot disk radius'' $R_{\m{h}}$. 
To put this in other words,   
the cooling front propagating 
inwards is stopped at a radius $R_{\m{h}}$ by the 
increasing X-ray luminosity, and the outer disk beyond 
$R_{\m{h}}$ remains shielded and cold. 
The cooling front at $R=R_{\m{h}}$ cannot propagate inwards 
until the  irradiation temperatures at $R~ \la ~R_{\m{h}}$  
drop below the hydrogen ionization temperatures. 
We shall see from the model fits that  $R_{\m{h}}$ actually remains  
roughly constant during the 
early decay phase.
$R_{\m{h}}$ separates regions of different evolution:  
The hot matter inside $R_{\m{h}}$ depletes faster than the cold matter 
outside $R_{\m{h}}$, due to the high viscosities operating in the hot
state. 

The shielding is removed  naturally when the thickness $h$
at $R~ \la ~R_{\m{h}}$ decreases enough for the central
X-rays to illuminate a part of the cold disk at 
$R > R_{\m{h}}$.
If the X-ray irradiation is still high enough to trigger a
thermal-viscous instability at the outermost cold disk then the resultant
disk evolution can account for the observed properties of the secondary
maximum. The increasing illuminated disk area (because 
of the removal of the shielding of the outermost disk) leads 
{\it first} to the 
observed optical rise. This is followed by an enhanced rate of mass
inflow. This scenario can account for the observed X-ray 
delay by a few days, corresponding to the viscous time scale for matter
released from the outermost disk to accrete to the inner disk.   

In the numerical calculations, we simply remove the
shielding at a parametrized time $t_1$ and see that the consequent model
X-ray photon flux curve fits well to the data.
With our numerical calculations we present disk thickness $h(R)$ 
profiles, for different times of the main decay before the onset of the 
secondary maximum, 
calculated using the midplane temperatures of our 
1-D model. These results indicate that, in the 1-D model,  
the disk does evolve to remove the shielding.

\section{The Bump}  
 
Outburst light curves of the  BH SXTs A0620-00 and  GS/GRS 1124-68 
also show a bump 
$\sim 150-170$ days after the main maximum. 
After the bump, the X-ray and the 
optical light curves decrease steeply and the system returns to quiescence 
(Tanaka $\&$ Shibazaki 1996). 
Unlike the main outburst and the secondary maximum, the bump 
does not rise very rapidly. 
The rise time scale is about few weeks. 
The optical maximum precedes the X-ray maximum by about two 
weeks for A0620-00 (Kuulkers 1998). The onset of the bump 
occurs when the system is in the so called low-hard 
state. 

Following the secondary maximum, the disk is again made of 
a cold outer region beyond $R_{\m{h}}$, shielded by the hot, irradiated  
inner disk that prevails at $R~ \la ~R_{\m{h}}$. Like the secondary
maximum, 
the bump is caused by irradiation reaching the outer disk and enhancing 
mass transfer. The different morphology of the bump indicates that 
in this instance, irradiation reaches the outer disk by a mechanism 
different from that causing the secondary maximum.    
The increase of the X-ray hardness ratio starts a few weeks 
before the onset of the bump. 
The hard X-ray photons are generally believed to be the up-scattered 
photons from a hot corona around the inner disk. A hot corona could be
formed due to the thermal instabilities from the optically thin surface
layers of the inner disk (Shaviv $\&$ Wehrse 1986), and it could be
stabilized by the soft photons coming from the photosphere of the
underlying thin disk. The seed photons provide 
cooling for the hot corona by means of inverse Compton scattering. 
As the accretion into the inner geometrically thin disk decreases, 
the black-body temperature of the inner disk and hence the number 
of seed photons also decrease.  
The cooling becomes less effective and the 
temperature of the corona increases.   
We suggest that the consequent 
increase in the size of the corona increase the efficiency 
of the X-ray irradiation. 
A thermal viscous instability could be triggered when 
the corona reaches dimensions large enough to illuminate the outer 
disk beyond the radius $R_{\m{h}}$. 
This leads to an increase first in the
optical luminosity  and then in the X-ray luminosity.  
The gradual expansion-illumination-ionization 
sequence is followed by the slow rise of the bump because 
of reduced viscosity and longer viscous time scales in this regime
compared to the conditions that prevailed at the secondary maximum.

\section{The Numerical Model}

In the thin disk approximation with 
$\Omega \simeq \Omega_{\m{K}} = (G M /R^3)^{1/2}$, 
the mass conservation equation 
\be
R \frac{\d \Sigma}{\d t} + \frac{\d}{\d R} (R \Sigma V_R ) = 0 ,
\label{1}
\ee
the angular momentum conservation equation
\be
R \frac{\d}{\d t}( \Sigma R^2 \Omega ) + 
\frac{\d}{\d R} (R \Sigma V_R R^2 \Omega ) = \frac {1}{2 \pi} 
\frac{\d g}{\d R} ,
\label{2}
\ee
together with the expression for the torque
\be
g = 2 \pi R \nu \Sigma R^2 \left( \frac{\d \Omega}{\d R} \right) ,
\label{3}
\ee
give a non-linear diffusion equation for
the surface density
\be
\frac{\d \Sigma}{\d t} = \frac{3}{R} \frac{\d}{\d R}
\left[ R^{1/2} \frac{\d}{\d R} (\nu \Sigma R^{1/2}) \right].
\label{4}
\ee 
With the assumption that $T_{\m{c}}^{4} >> T_{\m{eff}}^{4}$ ,~ we can
write
\be  
\frac{4 \sigma}{3 \tau} T_{\m{c}}^{4} = \frac{9}{8} \nu \Sigma 
\frac{G M}{R^3} = \sigma T_{\m{eff}}^4
\label{5}
\ee 
where $T_{\m{c}}$ and $T_{\m{eff}}$ are disk midplane and effective 
temperatures respectively, $\tau \sim \kappa_{R} \Sigma$ is the
vertically integrated optical
depth, and $\nu$ is the kinematic viscosity. We take the density 
$\rho = \Sigma / 2 h$ where $h$ is the pressure scale height of the disk
and the temperature $T=T_{\m{c}}$ to estimate  
the Rosseland mean opacities $\kappa_{R}$ by using the opacity tables   
for population I stars with mixture $X=0.7$ and $Z=0.02$ 
(Alexander $\&$ Fergusson 1994, for $log~T \le 3.7$ and 
Iglesias $\&$ Rogers 1996,  for $log~T > 3.7$). 
For the viscosity we use the standard $\a$ prescription 
$\nu = \a c_{\m{s}} h$  and adopt the commonly used bimodal
$\a$ parameter. We set $\a = \a_{\m{h}}=0.1$  
and $\a = \a_{\m{c}} = 3.3\times 10^{-2}$ for the hot and cold stable
states 
respectively. We neglect the radiation pressure and take the 
local sound speed $c_{\m{s}} = k T_{\m{c}} / \mu m_{\m{p}}$ 
where $k$ is the Boltzmann constant, $m_{\m{p}}$ is the proton rest mass, 
and $\mu$ is the mean molecular weight. We set $\mu=0.63$ in the hot
regime, and $\mu=0.87$ in the cold regime. 
Eq.(\ref{5}) assumes that the local viscously dissipated energy is
instantaneously radiated 
from the disk photosphere in the form of black-body radiation. 

By setting $x=2 R^{1/2}$ and $ S =x \Sigma$, Eq.(\ref{4}) can be written
in a simple form
\be
\frac{\d S}{\d t} = \frac{12}{x^2} \frac{\d^2}{\d x^2} (\nu S) .
\label{6}    
\ee
We divide the disk into equally spaced 200 grid points in $x$. This means
that the grid spacing in $R$ space decreases with decreasing $R$. This is
preferred for a better spatial resolution of the inner disk. Although 
our time steps are small enough ($\sim$ few seconds) for a good time
resolution, 200 grid points in $x$ space provide a low space resolution. 
Increasing spatial resolution requires shorter time steps and
extremely long computation times to scan the disk parameters to obtain 
a good fit to the observed X-ray photon flux curve of the BH SXT 
GS/GRS 1124-68. We therefore restrict the computation to 200 grid points
in $x$ at present. 

Since the GINGA ASM photon flux data (1-20 keV) 
is not de-convolved data, we convolve our 
model photon flux curve with the 
detector response matrix. 
The observed 
data and the response matrix
is provided by S.Kitamoto (private communication).   
After Eq.(\ref{6}) is
numerically solved, we further divide the main spatial grid 
at each time step into 20 equally spaced 
grid points in $x$ to determine the 
effective temperature distribution and the corresponding X-ray photon
flux with a higher accuracy. 
We take the black hole mass 
$M_{1} =6 M_{\odot}$,  the distance $d=3$ kpc (McClintock et al. 1992, 
West 1991). We neglect the neutral hydrogen absorption. 
The inclination angle was left as free parameter; 
$cos~i =0.88$ was obtained from our fits.

\subsection{X-ray Irradiation}

The X-ray irradiation flux is given by 
\be
\sigma \Tirr^4= \frac{\eta \dot{M}_{\m{in}} c^2 (1-\epsilon)}
  {4 \pi R^2} \left( \frac{\Hirr}{R} \right)^n 
  \left(\frac{d~ ln \Hirr}{d~ ln R}-1\right)
\label{7}
\ee

(Shakura $\&$ Sunyaev 1973), where $\eta$ is the efficiency of
the conversion of the rest mass energy into X-rays, $\epsilon$
is the X-ray albedo of the disk face, $\sigma$ is the 
Stefan-Boltzmann constant, $\Hirr$ 
is the local pressure scale height of the disk which should be calculated 
including the effect of X-ray irradiation itself (Dubus et al. 1999), 
and $\Mdot_{\m{in}}$ is the inner accretion rate. Shakura $\&$ Sunyaev
(1973) 
estimated that $n=1$ for the neutron star systems, and $n=2$ for the 
black hole systems. King $\&$ Ritter (1998) pointed out that $n$ could be
taken as unity for also the black hole 
systems, assuming that the source of 
the X-ray irradiation is a hot corona around the inner disk rather than the
thin inner disk surface. For a point source, Eq.(\ref{7}) can be
rewritten as 
\be
\sigma \Tirr^4 = C \frac{\Mdot_{\m{in}} c^2}{4 \pi R^2}    
\label{8} 
\ee
where
\be
C= \eta (1-\epsilon) \frac{\Hirr}{R}
  \left(\frac{d~ ln \Hirr}{d~ ln R}-1\right)
\label{9}.
\ee
$C$ can vary in a large interval for different choices of the parameters 
especially due to the uncertainty on the X-ray albedo $\epsilon$. 
For a point source, estimates for $C$ are usually in the range 
$10^{-4}-10^{-3}$ (Tuchman et al. 1990, de Jong et al. 1996, Dubus et al
1999) In our numerical model, we use Eq.(\ref{8}) 
parametrizing $C$ to calculate the irradiation temperatures
(See Sects. 4.4 and 4.5)). 

Detailed vertical disk structure analyses show that the S-shaped
characteristic of the equilibrium curves disappear above 
$\Tirr \sim 10^4$ K and the disk can remain in a hot stable state 
(Tuchman et al. 1990, Dubus et al. 1999). In these analyses the X-rays are
assumed to be absorbed in a thin layer of the disk surface. With this
assumption, these results show that the radiative hot, optically thick 
disks are modified by
the X-ray irradiation in the regions near the disk surface, while the
regions near the disk midplane, where most of the matter is
located, could remain unaffected. That is, an efficient X-ray irradiation
can change the stability criteria of the hot radiative  disks by changing
the boundary conditions at the disk surface without modifying
the disk midplane conditions. Following these results, in our numerical
model which does not address the vertical disk structure we estimate the 
viscosities by using the disk midplane temperatures $T_{\m{c}}$. The
irradiation temperatures, on the other hand, are used to calculate the 
critical minimum and maximum surface densities at a given time at
each grid point, and thereby to adjust the $\a$ parametrization (either 
$\a_{\m{h}}=0.1$ or $\a_{\m{c}}=0.033$) throughout the disk.

\subsection{Critical surface densities}         

Various authors have obtained minimum and maximum
critical surface densities of the hot and cold stable states for
non irradiated and steady state disk accretion 
by performing numerical fits to the turning points of S-shaped equilibrium
curves. Although the critical surface densities are model dependent,  
there is a close agreement between these results 
(Shafter et al. 1986, Cannizzo et al. 1988, Ludwig et al. 1994, 
Hameury et al. 1998). 

Most recently Dubus et al. (2001) computed 
$70\times 120 \times190 \times9 $ vertical 
disk structures in $\Omega_{\m{K}} ,\Sigma, T_{\m{c}}, \Tirr$ with
parameter ranges
convenient for SXTs for both irradiated and non-irradiated disks. They
found
\be 
\S_{\m{max}} = (10.8-10.3 \xi) \a_{\m{c}}^{-0.84} M_{1}^{-0.37+0.1 \xi} 
R_{10}^{1.11-0.27 \xi} ~\m{g cm}^{-2}
\label{10}
\ee 
\be
\S_{\m{min}} = (8.3-7.1 \xi) \a_{\m{h}}^{-0.77} M_{1}^{-0.37}
R_{10}^{1.12-0.23 \xi} ~\m{g cm}^{-2}
\label{11}
\ee
where $\xi=(\Tirr/10^4 \m{K})^2$, $M_1$ is the mass of the compact object
in units of solar masses, and $R_{10}=(R/10^{10} \m{cm})$.
These results provide an 
estimation of the critical surface densities for the irradiated unsteady 
disks, as well as for non-irradiated disks ($\xi=0$). 
We adopt the critical surface densities given by the 
Eqs.(\ref{10} $\&$ \ref{11}) with $\a_{\m{h}} =0.1$ and 
$\a_{\m{c}} =0.033$.

\subsection{The initial mass distribution and the early rise phase} 

We start with a Gaussian initial mass distribution at the beginning of the
outburst phase
\be
\Sigma (R, t=0) = \Sigma_0 ~ exp \left[-\left(\frac{R-R_0}{\Delta R} 
\right)^2 \right] 
\label{12}
\ee
where $R_0=9.0\times10^{10}$ cm is the location of the center of the
Gaussian , $ \Delta R=2\times10^{10}$ cm, and $\Sigma_0$ is the 
surface density at the center of the Gaussian at time $t=0$. 
The critical surface density $\Sigma_{\m{max}}$ is found to be close to 
$\Sigma_{0}$ in our best fits. We iterate and choose a fit with 
$\Sigma_{0}\simeq \Sigma_{\m{max}}$. This critical 
$\Sigma_{\m{max}}$ is
calculated for the non-irradiated case, since the quiescent X-ray
luminosity is in operation when the instability is triggered. 
We obtain the best
fits for  $\Sigma_0 = 1.1\times10^{3}$ g cm$^{-2}$ with chosen 
$R_0$. Substituting $\a_{\m{c}} =0.033$, the black hole mass 
$M_{\m{x}} = 6 \Msun$ and $R_0 = 9 \times 10^{10}$ cm in  Eq.(\ref{10}), 
we find  
$\Sigma_{\m{max}}=1.1 \times 10^{3}$ g cm$^{-2}=\Sigma_0$ for the 
non-irradiated case ($\xi=0$). The calculation has several free
parameters. These are ($i$) the time $t_0$, when the X-ray irradiation
starts - we found $t_0=3$ d. ($ii$) $T_{\m{eff,min}}$ above which the
disk is 
taken to be in the hot state until $t=t_0$ (after $t=t_0$ the critical
surface densities (Eqs.(\ref{10}$\&$\ref{11})) for an irradiated disk are
used). ($iii$) The time $t_1$ at which the shielding is removed, leading
to the secondary maximum. ($i$v) The time 
$t_2$ when the irradiation efficiency increases leading to the bump.
  
We start with the hot state viscosities ($\a =\a_{\m{h}}$). 
Until $t=t_0=3 d$ the radial grid points having 
$\Teff > 10000$ K are kept in the hot state, and we set 
$\a =\a_{\m{c}}$      
for lower $\Teff$. 
We switch on the X-ray
irradiation at $t\simeq 3$ d, and the outermost model disk having 
$\a =\a_{\m{c}}$ at that moment is assumed to be shielded, 
and so remains in the cold state until $t=t_1$ at which we remove
the shielding. To fix the given
parameters (for the first three days), we follow the quality of the  
fits until the end of the overall decay phase for many trials. This is
because the amount of shielded matter, 
the position of $R_{\m{h}}$, and the relative strength of the hot
and cold state viscosities  at  $t=t_1$ 
largely affect the model curve evolution after $t=t_1$. In other 
words, the evolving disk should not only give a good fit for the main 
outburst, but also end up with proper conditions  at $t=t_1$ to reproduce
the observed secondary maximum just by removing the shielding.  
We neglect the tidal forces and let the matter expand freely up to 
$R_{\m{out}}\sim 2.0\times 10^{11}$ cm which is about the truncation
radius for the assumed parameters of GS/GRS 1124-68  
(Frank et al. 1992). We consider the matter passing beyond $R_{\m{out}}$
as lost from the system.            

\begin{figure*}
\centering
\vspace{8.0cm}
\includegraphics{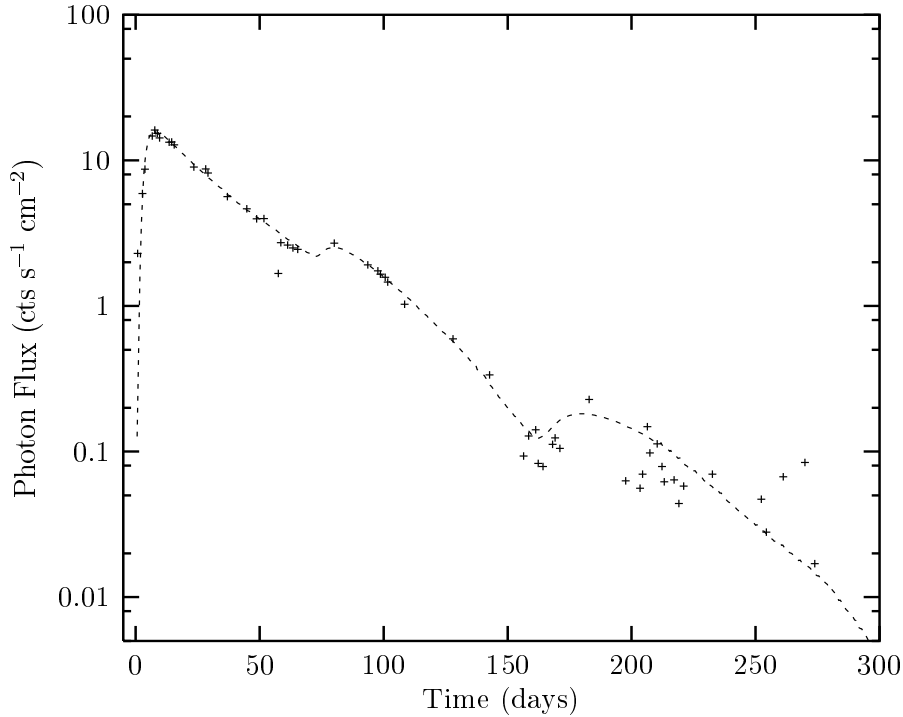}
\vspace{1.5 cm}
\caption{Observed GINGA ASM X-ray (1-20 keV) photon flux data and the
model curve (dashed line) with parameters presented in Sect. 4).}
\end{figure*}

\subsection{The main outburst and the secondary maximum}
  
From the time $t=3.0$ d to the onset of the secondary maximum 
($t=t_1$), the model disk evolves with a cold, shielded region 
outside the radius $R_{\m{h}}$ 
and an X-ray irradiated hot region inside $R_{\m{h}}$. For the X-ray
irradiated hot disk, the minimum critical surface densities and the 
X-ray irradiation temperatures are calculated 
by using Eq.(\ref{10}) and Eq.(\ref{8}) respectively. 
The parameter $C$ is determined from the fits. The model
X-ray photon flux curve seen in Fig. 1 is obtained with
$C=2.3\times 10^{-4}$ until the onset of the bump 
outburst after which we expect an increase in the strength of the 
X-ray irradiation (see Sect. 3). With this choice of $C$,
Fig. 3 shows that the hot disk radius $R_{\m{h}}$ 
remains constant from $t=3$~d to 
$t=t_1$. That is, the X-ray irradiation during this period is
strong enough to prevent the inward propagation of the cooling front. 
At $t=t_1$, we remove the shielding and apply the condition given by 
Eq.(\ref{10}) to determine the maximum critical surface densities 
of the outermost disk with the present X-ray irradiated conditions.       
After the removal of the
shielding, $R_{\m{h}}$ at first increases and then decreases gradually
governed
by the decreasing
strength of the X-ray irradiation (Fig. 3). 
A fraction of the outermost disk which
has so far remained shielded and cold, makes an upward transition and the
subsequent enhancement of the accretion rate from the outer disk to the
inner disk gives a model curve which fits well to the observed data. 
For the model curve presented in Fig. 1, we chose $t_1 \simeq 68$ days.
The surface density distribution one day before the triggering of the
secondary maximum is seen in Fig. 4. The removal of the shielding 
can be seen in the model disk scale height profile
evolution calculated by using the midplane temperatures.              
In Fig. 2 it is clearly seen that the shielding of the  outer
disk $(R > R_{\m{h}})$ which is present at the early phase of the main
decay is
removed before the onset of the secondary maximum ($t~ \la ~t_1$) for a
scattering region remaining roughly constant in
size ($l~ \la ~10^9$ cm) during the main decay. 
This result is obtained by using the midplane temperatures.
A full confirmation will invoke  2-D models calculating
the vertical disk structure including the effect of X-ray irradiation 
on the inner disk at $R < R_{\m{h}}$. 
This requires the extension of the 
work of Dubus et al. (2001) who calculated vertical structure for a 
fully irradiated  disk, for the case of a disk irradiated only 
at $R < R_{\m{h}}$ as in the present scenario.

\subsection{The Bump}

Our scenario to account for the bump is based on
the assumption that both the temperature and the effective scattering size
of the central hot corona increases when the decreasing  accretion rate of
the thin inner disk leads to a decrease 
in the soft photon supply that cools 
the corona. When the 
truncation of the inner disk starts,   
the inner disk radius $R_{\m{in}}$ changes in a way that is difficult to
estimate, due to uncertainties
in both the strength and the functional form of the evaporation. 
In the present work, we take  $R_{\m{in}}$ to be constant throughout the
entire outburst phase, 
assuming that the inner disk truncation does not start 
immediately upon the expansion of the corona. In other words,  during the
bump phase, the inner thin disk underlying the corona 
still extends down to the last stable orbit, but  it does not provide an
efficient cooling for the corona any longer. 
To implement this idea, we simply
increase the X-ray irradiation strength by a constant factor at 
$t=t_2$. This corresponds to an increase in the hot disk 
radius $R_{\m{h}}$ (Fig. 3), and 
leads to an enhancement in the mass transfer rate from the outer disk. 
For the model curve seen in Fig. 1, $t_2 = 155$~d, and the irradiation
strength is increased by a factor 1.6 at $t=t_2$. 
The surface density distribution one day before the triggering of the 
bump is presented in Fig. 5. 

\begin{figure}
\vspace{7.0cm}
\includegraphics{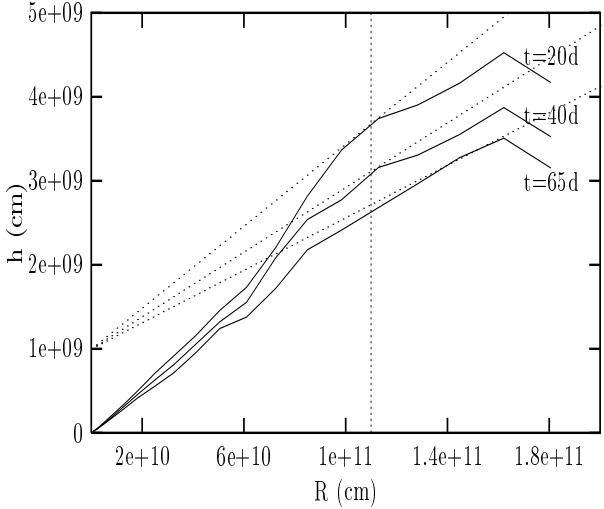}
\vspace{0.1 cm}
\caption{Pressure scale height $h$ profiles calculated by using the
midplane temperatures given by our numerical model
at three different times before the onset of the secondary maximum
($t \sim 68$ d). The local fluctuations are smoothed with the same binning
size for all the given curves to clarify the evolution. The vertical
dotted line corresponds to $R_{\m{h}}$ in our model which remains constant
during
the main decay.
The model disk evolves in a trend to remove the shielding
at about the observed secondary maximum for an inner scattering region
with a size $~ \la ~10^9$ cm. See the text for further
explanation.} 
\end{figure}

The fluctuations in  
the observed X-ray photon flux curve of the source GS/GRS 1124-68 during
the bump have small
time scales, sometimes as small as a day. 
Variations in the inner disk 
radius due to the truncation of part 
of the inner disk and the consequent refilling 
in response to newly arising  
density gradients could be a possible reason for these fluctuations. 

\begin{figure}
\vspace{7.0cm}
\includegraphics{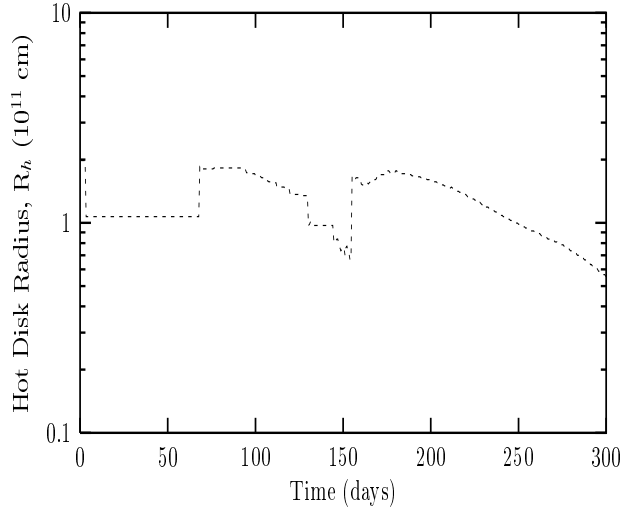}
\vspace{0.1 cm}   
\caption{Evolution of the hot disk radius $R_{\m{h}}$ corresponding to the
model curve seen in Fig(1). The triggering times are $t_1 =68~d$ for the
secondary maximum and $t_2 =155$~d for the bump.
The time intervals between the trigger and the  
starting of the rise in the X-ray flux are $\sim 5$~d and
$\sim 10$~d for the secondary maximum  and the bump
respectively.}
\end{figure}  

\begin{figure} 
\vspace{7.0cm}
\includegraphics{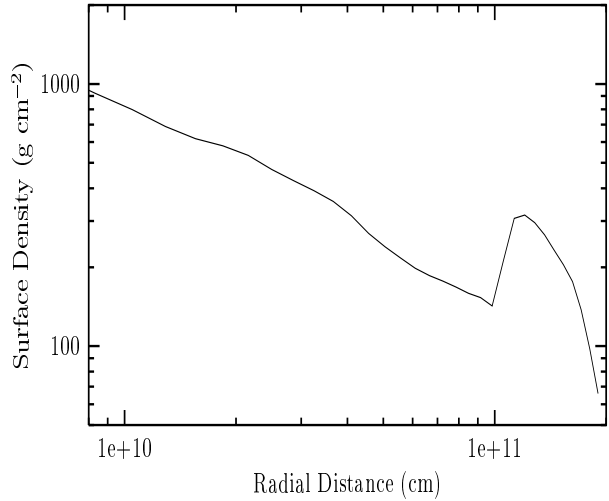}
\vspace{0.1 cm}
\caption{Radial surface density profile one day before the onset of the
secondary maximum ($t=67$~d). $R_{\m{h}} \sim 1\times 10^{11}$ cm.}
\end{figure}

\begin{figure}
\vspace{7.0cm}
\includegraphics{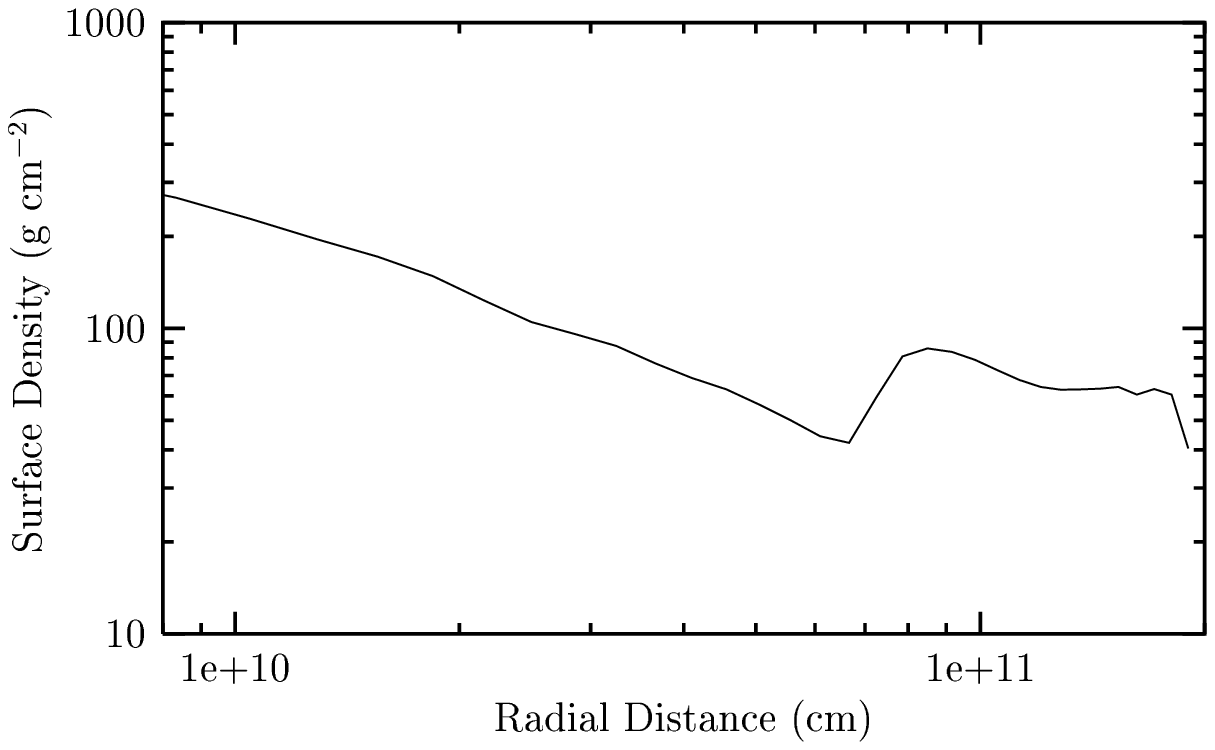}
\vspace{0.1 cm}
\caption{Radial surface density profile one day before the onset of the
bump ($t=154$~d). $R_{\m{h}} \sim 7\times 10^{10}$ cm.}
\end{figure}

\subsection{The simplifications and the reliability of the numerical
model}

In 2-D irradiated disk  models the transitions between the hot and 
the cold stable states are determined by the local disk temperatures 
calculated by the vertical disk analyses  considering 
the irradiation flux through the disk surface. 
The critical irradiation temperatures are different for 
different surface densities. This becomes important in modeling the 
unsteady and irradiated disks. Since the irradiation temperatures 
are mainly determined by the conditions of the inner disk, 
in a 1-D disk model it is not possible to determine the critical irradiation 
temperatures for varying surface densities at a particular radial distance 
$R$. To choose a uniform critical irradiation temperature 
$T_{\m{irr}}\sim 10^4$ K may lead to erroneous results. 
For these reasons, in our 1-D model we use the critical surface densities 
for irradiated and unsteady disks calculated by Dubus et al.(2001) (DHL). 

We checked the run of our numerical code using the presented 
results of the detailed 2-D model of DHL. 
The DHL model assumes a completely irradiated disk (no shielding) whereas 
in our numerical model for GS/GRS 1124-68 the outer disk 
($R > R_{\m{h}}$) 
is assumed to be shielded during the main decay. For comparison, 
in an illustrative model we do not include the shielding and adjust the      
amplitude of our initial Gaussian
mass distribution so that the $\Mdot_{\m{in}}$ peaks at the same level as 
the DHL curve. We plot this curve together with the points estimated from
Fig. 6 of DHL in our Fig. 7. We also present the 
surface density   profiles at different times of the  
$\Mdot_{\m{in}}$ curve in Fig. 8.  The
irradiation temperatures are similar in both models. 
We set $\a_{\m{h}}=0.1$ in
our illustrative  model, while $\a_{\m{h}}=0.2$ in the DHL model. 
This is reasonable, 
since we employ the midplane temperatures $T_{\m{c}}$ in calculating the
viscosities 
whereas in a 2-D model effective viscosities correspond to a temperature 
that is between $T_{\m{c}}$ and $\sim \Tirr$ (or $\Teff$). The second
difference
is that the outer disk radius is determined by the tidal forces in the DHL
model, whereas  it is fixed  and the matter going beyond this radius is
assumed   
to be lost from the system in our illustrative model. 
The $\Mdot_{\m{in}}$ and the surface density evolutions given by our 
illustrative model are seen to be similar to those given by DHL 
(our Fig. 7 $\&$ Fig. 8, and Fig. 6 of DHL).    

\begin{figure}
\vspace{4.2cm}
\includegraphics{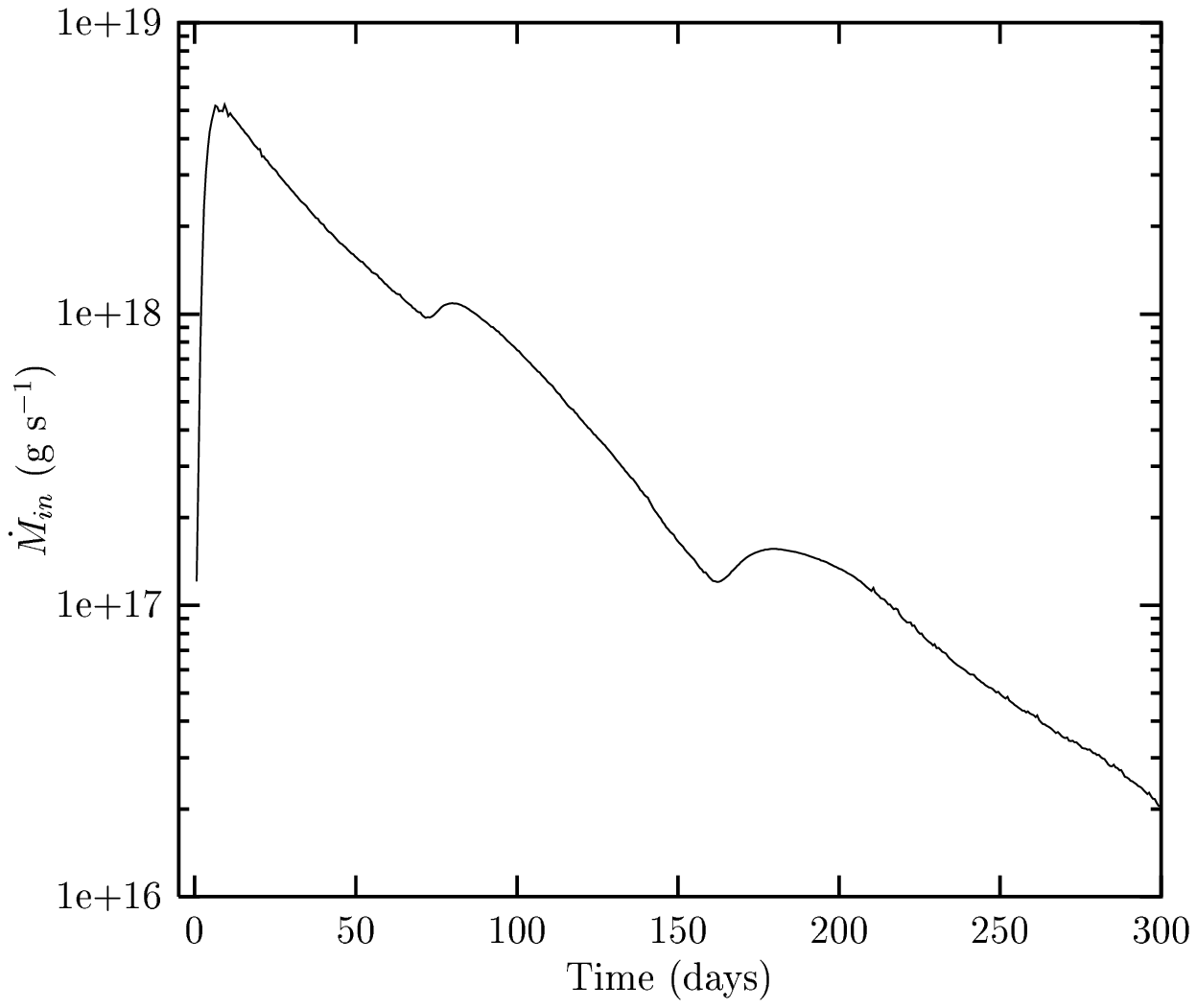}
\vspace{3.5 cm}   
\caption{Inner disk accretion rate $\Mdot_{\m{in}}$ evolution for the
model parameters presented in Sect. 4.}
\end{figure}

\begin{figure}
\vspace{6.8cm}
\includegraphics{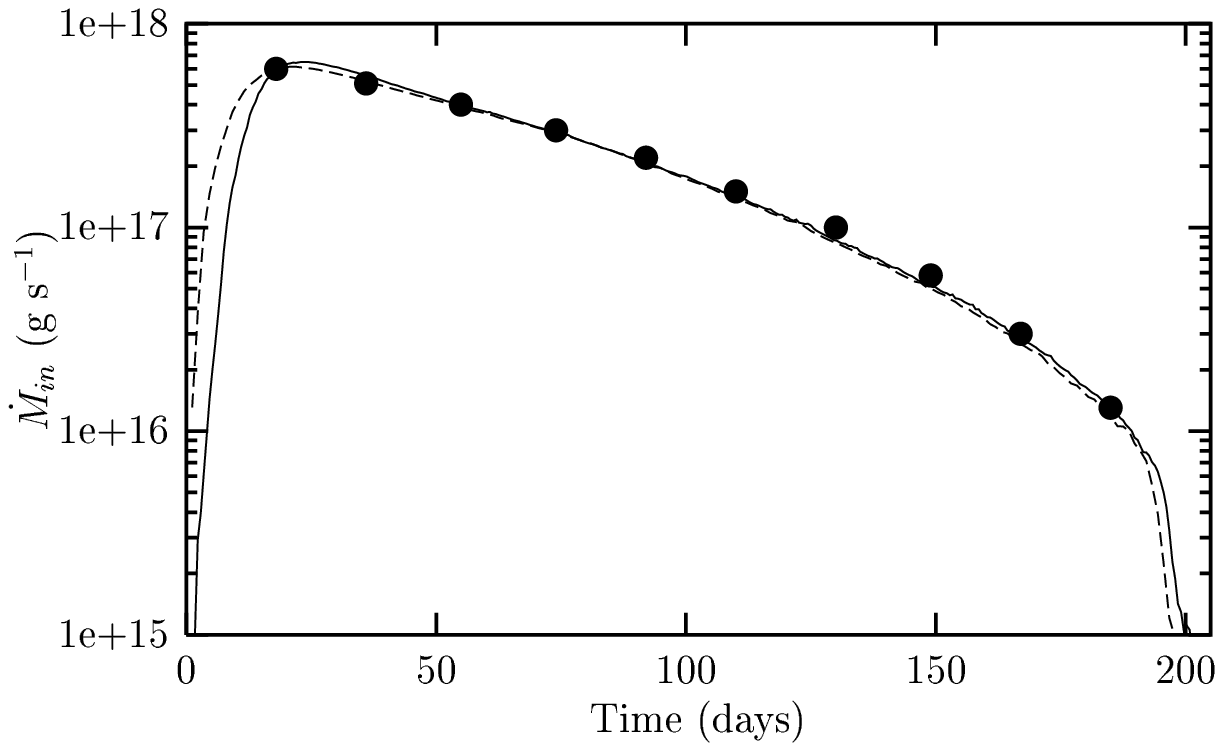}
\vspace{0.1 cm}   
\caption{Inner accretion rate evolution for the illustrative
$\Sigma \propto R$ (solid curve) and Gaussian (dashed curve)
initial mass distributions with the same disk parameters. The
filled circles are estimated from Fig. 6 of Dubus et al.(2001) (DHL)
for comparison. The amplitudes of the initial surface densities
were adjusted to match our model maxima to the maximum of
the DHL model. The shielding
is not included in our illustrative models.}
\end{figure}
In Fig. 7 we present another $\Mdot_{\m{in}}$ curve for a
different initial mass distribution ($\Sigma\propto R$) without changing
the other parameters. This is to show
that the different initial mass distributions 
give similar  $\Mdot_{\m{in}}$
curves, as long as most of the matter is initially located at the
outer disk.      

\begin{figure}
\vspace{7.0cm}
\includegraphics{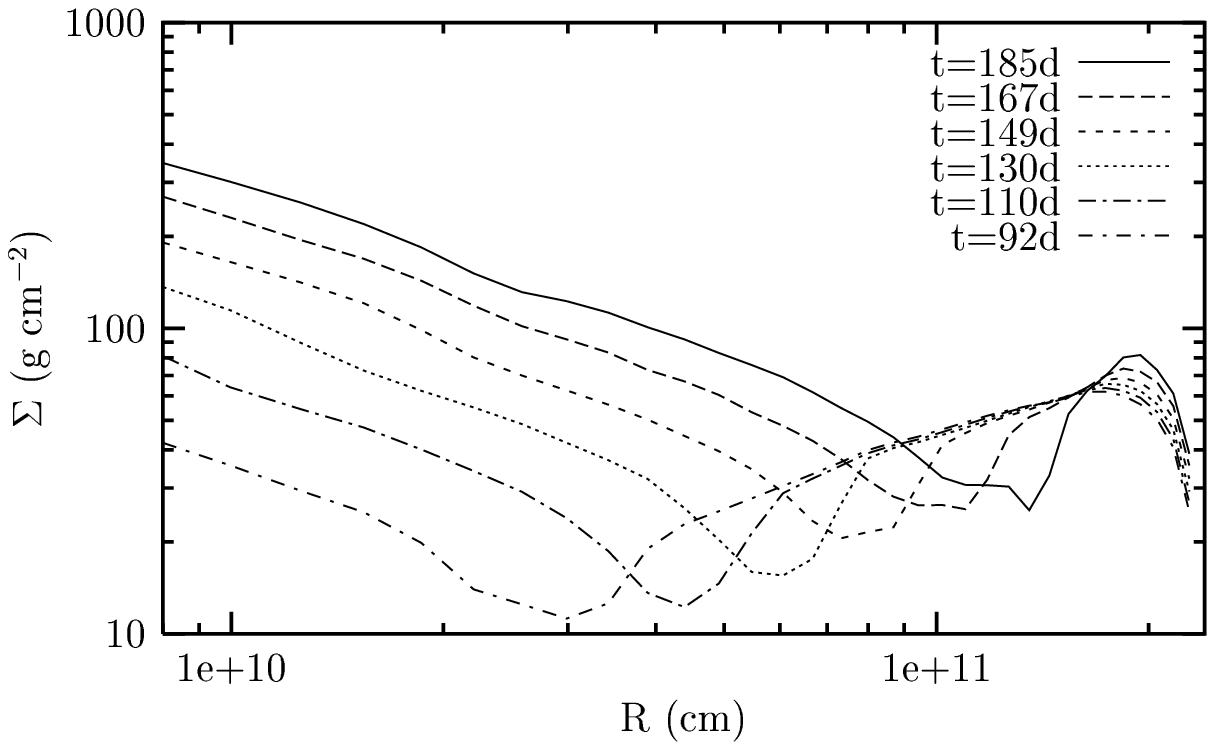}
\vspace{0.1 cm}
\caption{ Surface density distributions corresponding to different points
on the $\Mdot_{\m{in}}$ curve shown in Fig. 7 (dashed curve)   
given by the illustrative Gaussian initial mass distribution. These
are seen to be similar to those presented in Fig. 6 of Dubus
et al.(2001) for the same inner accretion rates.}
\end{figure}

\section{Discussion and Conclusions}

In general, soft X-ray transients (SXTs) show a variety of outburst light
curves (Chen et al. 1997). In the present work we concentrate on the
outburst light curves having a fast rise and a long lasting 
($\sim 250$ days) decay behavior exhibited by the  
BH SXTs A0620-00, GS/GRS~1124-68, GS~2000+25 and GRO~J0422+32. 

We presented a new scenario to explain these outburst light curves 
together with the  characteristic secondary maxima and the bumps    
seen on their decay phases. The explanations 
are based on the disk instability models including the effect of 
X-ray irradiation, and can be summarized as follows: 
At the beginning of the main outburst, 
the accumulated cold matter 
is released by a thermal-viscous instability. This matter expands 
to both lower and larger radii. Part of the
outermost disk that makes a transition back to the cold state
before the efficient X-ray irradiation has started, 
initially remains shielded by the hot inner disk. 
The surface densities of the hot inner disk decrease faster 
than the surface densities  of the shielded  disk at $R>R_{\m{h}}$.
Finally, the shielding is removed after the pressure scale 
height of the disk at $R~ \la ~R_{\m{h}}$~ has 
decreased enough for the central
X-rays to illuminate the cold disk at  $R > R_{\m{h}}$. 
The consequently triggered instability 
outside $R_{\m{h}}$ leads to enhanced mass flow resulting in the  
secondary maximum. The hot disk radius $R_{\m{h}}$ 
first increases with the
triggering, and later gradually decreases, governed by the
decreasing irradiation flux through the outer disk. 
When the inner accretion rate 
$\Mdot_{\m{in}}$ becomes comparable to the evaporation rate, 
the temperature of the 
corona  increases due to decreasing cooling rate. Consequently, 
the corona which is assumed to be the source of indirect illumination 
heats up and expands leading to more efficient X-ray irradiation 
and a new thermal-viscous instability 
beyond the present position of $R_{\m{h}}$. 
The subsequent enhancement of the mass transfer rate from the outer disk 
results in the observed bump.
     
We tested this scenario by using a one dimensional numerical model 
(Sect. 4). The model giving the X-ray photon flux curve given in
Fig. 1 works briefly as follows:  At the beginning
of the main outburst, the accumulated matter
is represented by a Gaussian mass
distribution at the outer disk. The center of the Gaussian is taken
to be $R=R_0 =9\times 10^{10}$ cm.   
Its maximum density  $\Sigma_{0}=1.1\times 10^3$ g cm$^{-2}$ is
comparable to the 
maximum critical surface density of the cold state for the chosen 
parameters $M_1=6\Msun,~ \a_{\m{c}}=0.033 ~
and~ R_0 =9\times 10^{10}$ cm
(Cannizzo et al 1988, Dubus et al 2001). 
We start with the hot state viscosities 
($\a=\a_{\m{h}}=0.1$). Until the X-ray irradiation is switched on at
$t=t_0$~d, 
we set $\a=\a_{\m{c}}$ for the grid 
points with the effective temperatures 
$\Teff$ decreasing below 10000 K, and the grid points having higher
$\Teff$ are kept in the hot state. At $t=t_0=3.0$~d, the outermost
cold disk 
region with $\a=\a_{\m{c}}$ ($R>R_{\m{h}}$) is taken to be shielded. 
After $t=t_{0}$ d, we use the critical surface
densities obtained for irradiated and unsteady disks (Dubus et al. 2001). 
 From $t=t_0$ to $t=t_1$, the disk evolves with a hot inner region  
($R < R_{\m{h}}$) and a cold shielded outer region  ($R > R_{\m{h}}$). 
The hot disk radius $R_{\m{h}}$ is 
free to move, e.g. inwards, if the surface
densities at the grid 
points inside $R_{\m{h}}$ decrease below the critical minimum surface
densities
of the present irradiated conditions. We find that  $R_{\m{h}}$
remains constant until the onset of the secondary maximum ($t=t_1$) for
the irradiation strength ($C=2.3\times 10^{-4}$) that gives the best fit
to the data until the onset of the bump ($t=t_2$).     
Our 1-D numerical model does not give the irradiated vertical 
disk structure, 
so we remove the shielding at a  parametrized time $t=t_1$ 
whose value, $t_1 =68$ d, is determined from the best fits. 
We assume that all the disk has become irradiated 
at $t=t_1$, and the part of the outer disk which has so far remained
shielded and cold makes a transition to the hot state at $t=t_1$. 
After $t=t_1$, $R_{\m{h}}$ first increases by the removal of the shielding 
and then decreases gradually governed by the decreasing irradiation
strength. To check 
for an indication that the shielding is removed in the 1-D model, we  
plot the disk thickness profiles calculated 
by the midplane temperatures 
at different times of the main decay phase. We see that 
the shielding of the outer disk ($R >R_{\m{h}}$) which is present 
during the main decay is removed at $t \sim t_1$  
for an inner scattering region 
with a size $l~ \la ~10^9$ cm. The confirmation 
of the initial settling and the removal of the shielding 
in a self-consistent model will be the subject of future work.  
The removal of the shielding at $t=t_1$ exposes the outer disk to
irradiation. This leads to the secondary maximum, first in the optical,
and then in the X-rays, as the enhanced mass flow reaches the inner disk.
With $t_1 =68$~d, the model reproduces the secondary maximum.

Increasing irradiation efficiency through an expanding corona, 
rather than removal of shielding,   
is taken to be the cause of the 
bump. At $t=t_2=155$ d,  the irradiation temperatures are increased 
by a constant ($\sim 1.6$) factor chosen to fit the amplitude of the 
bump. In our model, the transitions to the hot state  
at $t=t_1$ and later at $t=t_2$ are taken to occur simultaneously  
in the newly irradiated regions,  
rather than by the propagation of thermal fronts. 
The tidal forces at outer disk and the accretion
from the companion are neglected. 
We obtained the outer disk radius 
$R_{\m{out}}=2\times 10^{11}$ cm from our fits. 
This is about the  truncation radius 
$R_{\m{tr}}\simeq 0.9 R_{\m{L}_{1}}$. 
The matter going beyond this radius
is assumed to be lost from the system.
In Sec. 4.6 we showed that the simplifications of our  
numerical model does not significantly affect the results. 

The maximum of the outburst in X-rays is reached at 
$\sim 7$ d in the model. 
The time interval between the triggering and the beginning of the rise
given by the model is
about 5 days for the secondary maximum and 10 days for the bump. 
These results are in good agreement  with
the reported delays of the X-ray maxima with respect to the optical 
which are 
$\sim 4-6$ days for the main outburst and the secondary maximum, and 
about two weeks for the bump 
(Kuulkers 1998, Ebisawa et al. 1994, Orosz et al. 1997).

We have thus shown, through a numerical 
model with plausible parameters (Fig. 1), 
that the rise and the decay of BH SXT outbursts including the 
secondary maximum and the bump 
are a simple interactive history of the effects of 
viscous diffusion, irradiation and hot-cold state transitions. Both the
secondary maximum  and the bump are events taking place,
only once, in the evolution of the disk after the main outburst.      

The observed
fluctuations shown during the bump could be due to the
small variations of the inner disk radius when the evaporation rate
becomes comparable to the inner disk accretion rate. The evaporation is
expected to
be strongest at the inner disk and steeply decrease with increasing
radius (Meyer et al. 2000, Dubus et al 2001). Then the large density
gradients due to an evacuation of the inner disk by the evaporation
could result in the variations of the inner accretion rate, which could
easily modify the X-ray flux. In our numerical model, we do not address
these fluctuations. A reasonable fit to what might plausibly be the mean
behavior of the
bump is produced for GRS/GS 1124-68.

In our numerical model we do not include possible mass losses from
the disk surface due to evaporation (Meyer et al. 2000, 
Dubus et al. 2001, Shaviv et al. 1999). The X-ray irradiation is 
expected to increase the wind losses from the outer disk 
(de Kool $\&$ Wickramasinghe 1999). Although there is no general
agreement on the physics of evaporation, it has probably no
significant effect on the outburst light curves (Dubus et al. 2001).
However, there is a  consensus that evaporation is important  
in the quiescent states, and the inner disk is probably truncated 
because of the resultant mass losses. 
Meyer-Hofmeister $\&$ Meyer (1999) and Dubus et al. (2001) obtain the 
long outburst recurrence times of SXTs ($\sim$ few ten years or more)  by
including evaporation in their numerical model.

~~~

{\it \bf Acknowledgments}
We acknowledge support from the BDP Doctoral Research program of
 T{\"U}B{\.I}TAK (The Scientific and Technical Research Council of 
Turkey) and through The High Energy Astrophysics Research Group 
TBAG-\c{C}-4 of T{\"U}B{\.I}TAK. We thank S. Kitamoto for providing 
GINGA ASM outburst data for the source GS/GRS 1124-68.
Part of this research was done during 
the Summer Research Semester in Astrophysics (2001) of 
the Feza G{\"u}rsey Institute of T{\"U}B{\.I}TAK. 
MAA acknowledges support from the Turkish 
Academy of Sciences. We thank the referee for useful criticism.

\end{document}